
\input phyzzx
%
%
\def\np{Nucl. Phys. }
\def\pl{Phys. Lett. }
\def\prl{Phys. Rev. Lett. }
\def\pr{Phys. Rev. }

\def\cmp{Comm. on Math. Phys. }

\def\mpl{Mod. Phys. Lett. }
\REF\DSL{E. Brezin and V.~A. Kazakov \pl {\bf B236} (1990) 144; \nextline
M.~Douglas and S.~Shenker, \np {\bf B335} (1990) 635; \nextline
D.~J.~Gross and A.~A.~Migdal \prl {\bf 64} (1990) 127.}
\REF\GM{D. J. Gross and A. A. Migdal, \np {\bf B340} (1990) 333.}
\REF\KdV{T.~Banks, M.~Douglas, N.~Seiberg and S.~Shenker
\pl {\bf B238} (1990) 279.}
\REF\KAZ{V.~A.~Kazakov, \mpl {\bf A4} (1989) 2125.}
\REF\BREZ{E. Brezin, E. Marinari and G. Parisi, \pl {\bf B242} (1990) 35.}
\REF\LYK{S.~Chaudhuri and J.~D.~Lykken, {\it Analyzing the Solutions of
Hermitian Matrix Models}, Fermilab Preprint FERMILAB--PUB--90/267--T.}
\REF\DAVA{F. David, \mpl {\bf A5} (1990) 1019.}
\REF\FLOW{M. R. Douglas, N. Seiberg and S. H. Shenker, \pl {\bf B244}
(1990) 381.}
\REF\ANALG{G.~Gibbons, S.~Hawking and M.~Perry
\np {\bf B138} (1978) 141; \nextline
K.~Gawedzki and A.~Kupiainen, \np {\bf B257[FS14]} (1985) 474.}
\REF\DAVB{F.~David, \np {\bf B348} (1991) 507, and {\it
Non-Perturbative Effects in Two Dimensional Gravity}, in ``Two
Dimensional Gravity and Random Surfaces'', Proceedings of the
$8^{\rm th}$
Jerusalem Winter School for Theoretical Physics (1990/91).}
\REF\SILVA{P. G. Silvestrov and A. S. Yelkhovsky, \pl {\bf B251} (1990)
525.}
\REF\GH{J. Greensite and M. B. Halpern, \np {\bf B242} (1984) 167.}
\REF\PAR{E. Marinari and G. Parisi, \pl {\bf B240} (1990) 375.}
\REF\QM{J. Ambj\o rn, J. Greensite and S.~Varsted, \pl {\bf B249} (1990)
411; \nextline
J. L. Miramontes, J. S\'anchez Guill\'en and M.~A.~H. Vozmediano, \pl
{\bf B253} (1991) 38.}
\REF\KM{M. Karliner and A. A. Migdal, \mpl {\bf A5} (1990) 2565.}
\REF\SUSY{J.~Gonz\'alez and M.~A.~H.~Vozmediano
\pl {\bf B247} (1990) 267; \nextline
J.~Gonz\'alez, \pl {\bf B255} (1991) 367; \nextline
A.~Dabholkar, {\it Fermions and Non-Perturbative Supersymmetry
Breaking in the One Dimensional Superstring},
Rutgers Univ. Preprint RU--91/20.}
\REF\AMBA{J. Ambj\o rn and J. Greensite, \pl {\bf B254} (1991) 66.}
\REF\GZJ{P. Ginsparg and J. Zinn--Justin, \pl {\bf B255} (1991) 189,
and {\it Action Principle and Large Order Behaviour of
Non-Perturbative Gravity},
in ``Random Surfaces, Quantum Gravity and Strings (Carg\`ese 1990)'',
Ed. by O.~Alvarez, P.~Windey and E.~Marinari (to appear).}
\REF\MORRIS{S. Dalley, C. Johnson and T. Morris, {\it Non-Perturbative
Two-Dimensional Quantum Gravity}, Univ. of Southampton Preprint
SHEP~90/91--28, and {\it Non-Perturbative Two-Dimensional Quantum Gravity,
Again}, Univ. of Southampton Preprint SHEP 90/91--35, in
``Proceedings of the
Workshop on Random Surfaces and 2D Quantum Gravity'', Barcelona 1991,
to appear in Nucl. Phys. {\bf B} (Proc. Suppl).}
\REF\MORAMB{J. Ambj\o rn, C. V. Johnson and T. R. Morris, {\it Stochastic
Quantization vs. KdV Flows in 2D Quantum Gravity}, Univ. of Southampton
and Niels Bohr Inst. Preprint SHEP~90/91--29, NBI--HE--91--27.}
\REF\MOORE{G. Moore, M. Ronen Plesser and S. Ramgoolam, {\it Exact
S-Matrix for 2D String Theory}, Yale Univ. Preprint YCTP--P35--91.}
\REF\BIPZ{E. Brezin, C. Itzykson, G. Parisi and J.~B.~Zuber, \cmp {\bf 59}
(1978) 35.}
\REF\Demeterfi{K.~Demeterfi, N.~Deo, S.~Jain and C.~Tan
\pr {\bf D42} (1990) 4105; \nextline
J. Jurkiewitz, \pl {\bf B261} (1991) 260.}
\REF\SILVB{P. G. Silvestrov, {\it Two Dimensional Gravity from d=0 and d=1
Matrix Model}, Novosibirsk Inst. of Nucl. Phys. Preprint 91--50.}
\REF\Geli{J.~Gonz\'alez and M.~A.~H.~Vozmediano
\pl {\bf B258} (1991) 55; \nextline
O.~Diego and J.~Gonz\'alez, {
\it Multicriticality in the Stabilized 2D Quantum Gravity},
Inst. de Estructura de la Materia Preprint IEM--FT--43/91.}
\REF\SHEN{S.~Shenker, {\it The Srength of Non-Perturbative Effects in
String Theory},
in ``Random Surfaces, Quantum Gravity and Strings (Carg\`ese 1990)'',
Ed. by O.~Alvarez, P.~Windey and E.~Marinari (to appear).}
\REF\BESSIS{D. Bessis, C. Itzykson and J.~B.~Zuber
Adv. in Appl. Math. {\bf 1} (1980) 109.}
\REF\BARNA{J. L. Miramontes and J. S\'anchez Guill\'en, {\it Instantons in
the Quantum Mechanical Framework of 2D Gravity}, in ``Proceedings of the
Workshop on Random Surfaces and 2D Quantum Gravity'', Barcelona 1991,
to appear in Nucl. Phys. { \bf B} (Proc. Suppl.).}
\REF\COL{S. Coleman, \pr {\bf D15} (1977) 2929; \nextline
C. G. Callan and S. Coleman, \pr {\bf D16} (1977) 1762.}
\REF\ZJBOOK{J. Zinn--Justin, {\it Quantum Field Theory and Critical
Phenomena}, Clarendon Press, Oxford 1990.}
\REF\PETRO{C. Bachas and P. Petropoulos, \pl {\bf B247} (1990) 363.}
\REF\TIM{T. Hollowood, J. L. Miramontes, A. Pasquinucci
and C. Nappi, {\it Hermitian vs.
Anti-Hermitian 1-Matrix Models and their Hierarchies},
Princeton Univ. and Inst. for Adv. Study Preprint PUPT--1280,
IASSNS--HEP--91/59, to appear in \np {\bf B}.}
\nonstopmode
\overfullrule = 1pt
\normalbaselineskip  = 20pt plus 0.2pt minus 0.1pt     
\normaldisplayskip   = 11pt plus 4pt minus 5pt         
\normaldispshortskip = 5pt plus 5pt                    
\nopubblock
\rightline{CERN--TH. 6323/91}
\rightline{November 1991}
\titlepage
\title{\bf Universality and Non-Perturbative Definitions of 2D
Quantum Gravity from Matrix Models}
\author{{\bf J.Luis Miramontes}\footnote{*}{e-mail:
miramont@cernvm.cern.ch}}
\address{\it Theory Division, CERN
  \break CH--1211 Gen\`eve 23, Switzerland}
\andauthor{{\bf Joaqu\'\i n S\' anchez Guill\' en}\footnote{**}{Permanent
address: Dept. of Particle Physics, Univ. of Santiago, E--15706
Santiago de Compostela (Spain).
e-mail: joaquin@euscvx.decnet.cern.ch}}
\address{\it Department of Theoretical Physics
\break University of Lund
\break Solvegatan 14A
\break S--22362 Lund, Sweden}
\abstract{The universality of the non-perturbative definition of
Hermitian one-matrix models following the quantum, stochastic, or
$d=1$-like stabilization is discussed in comparison with other
procedures. We also present another alternative definition, which
illustrates the need of new physical input for $d=0$ matrix models
to make contact with 2D quantum gravity at the non-perturbative level.}
\endpage
\pagenumber=1
\pageno=1
\normalbaselineskip  = 20pt plus 0.2pt minus 0.1pt      
\normaldisplayskip   = 11pt plus 4pt minus 5pt          
\normaldispshortskip = 5pt plus 5pt                     
\hoffset =  0.0 in \hsize = 6.5 in                      
\voffset = -0.1 in \vsize = 9.0 in                      
\normalspace
\chapter{INTRODUCTION}

The double-scaling limit of the matrix models
provides a regularized version
of the sum over arbitrary surfaces and, consequently, of 2D quantum
gravity$^{\DSL,\GM}$. In the
case of the Hermitian one-matrix models the result, by
now well-known, is that there is an infinite number of local operators,
$\bf \hat O_k$, $k=0,1,\ldots$, which
realize a KdV flow structure$^{\GM,\KdV}$, and whose simplest
operator is the puncture
$O_0\equiv {\bf \hat P}$. Moreover, the specific heat of
the model satisfies a
differential equation known as the ``string equation'', whose scaling
solutions in the cosmological constant are the multicritical
models$^{\KAZ}$. Introducing the couplings of the operators
$\bf \hat O_k$, $t_k$, the string equation is
$$
\sum_{k=1}^{\infty} (2k+1) t_k R_k [f] = -T,
\eqn\stringeq
$$
where $T\equiv t_0$ is the renormalized
cosmological constant, $f= d^2 \ln{\bf Z}/ d
T^2 \equiv \langle {\bf \hat P \hat P}\rangle$ is the specific heat, $R_k$
are the Gelfand--Dikii polynomials that define the KdV flows
$$
{\partial  f\over\partial t_k} \equiv \langle {\bf \hat P \hat P} {\bf
\hat O_k} \rangle =
R_{k}'
\eqn\KdV
$$
and primes denote derivatives with respect to $T$.
The $k^{\rm th}$ multicritical model, which is defined
by the asymptotic behaviour
$f\approx T^{1/k}+\ldots$ for large positive values of $T$,
corresponds to $t_j=(2k+1)^{-1}\delta_{j,k}$.
The simplest one realized in the matrix model is
pure gravity ($k=2$), whose string equation is the Painlev\'e I equation
$$
f^2 - {1\over3} f'' = T .
\eqn\Painleve
$$

In any of these models, the contribution of the surfaces with an arbitrary
fixed genus
to the specific heat is obtained by expanding the solution of the string
equation in powers of the string
coupling constant, which is related to the cosmological constant,
$g_{str}\propto T^{-{2k+1\over2k}}$. It was also expected
that the string equation, which, strictly speaking, is only valid order by
order in the genus expansion,
would define 2D quantum gravity at the non-perturbative level as well.
This would require to fix the string equation's boundary
conditions. Nevertheless, although it is possible for the odd-$k$
models$^{\BREZ,\LYK}$, the even-$k$ ones
present additional complications. In the particular case of pure
gravity, it has
been argued that no real solution of the Painlev\'e~I equation can
be compatible
with the loop equations (or Schwinger--Dyson equations) of the matrix
model$^{\DAVA}$. Furthermore, the behaviour of the (KdV) flow between the
(well defined) $k=3$ and $k=2$ models indicates the non-perturbative
instability of pure gravity$^{\FLOW}$.

The definition of the models with even-$k$ involves matrix models whose
potentials are not bounded from below, and this seems to be the origin of
the problems$^{\BREZ}$. Consequently,
the two standard procedures to stabilize field theories with unbounded
actions have been tried in this context.
The first method is the analytic continuation of the
dominant term of the potential from positive
to negative sign, while a simultaneous deformation of
the contour of integration on the matrix eigenvalues is
performed$^{\ANALG}$.
Unfortunately, this
method leads to a non-perturbative imaginary part in the specific heat,
which makes the result unphysical$^{\DAVA,\DAVB,\SILVA}$.
Using the second generic method, inspired in stochastic quantization and
proposed in general by Greensite and Halpern$^{\GH}$, the ill-defined
($d=0$) matrix model is formulated like
a well-defined ($d=1$) quantum mechanical
system$^{\PAR,\QM,\KM}$.
Following this method, the specific heat remains real, in contrast with
the result of the analytic continuation.\footnote{a}{In this paper,
we shall not be interested in the supersymmetric aspects of the
original proposal of Marinari and Parisi$^{\PAR,\SUSY}$.}

In addition to its relevance in the context of the matrix models, the
study of the stochastic-like or quantum mechanical method
is interesting by itself, as a non-trivial example of the
general stabilization mechanism proposed in ref.[\GH].
So far, the universality
of this regularization has not been fully understood.
In fact, all the previous papers
concentrate on the definition of pure gravity using the
simplest potentials of degree 3
or 4, either in the WKB approximation$^{\PAR,\QM}$, or in numerical
computations$^{\KM,\AMBA}$. The main reason for this lack of generality
is that almost all the calculations rely on the fact that the quantum
mechanical system is an ideal Fermi gas for the simplest potentials
of degree less than $4$. In the next section 2, we
show that it is possible to recover the ideal Fermi gas
picture for an arbitrary potential through the use of a mean field
approximation (\` a la Hartree--Fock), which agrees with the semiclassical
limit of the matrix model
given by the loop equations. This result ensures the
universality of this regularization procedure for any given multicritical
model
(not only pure gravity), and provides a connection between the quantum
(Fokker--Planck) potential and the semiclassical density of eigenvalues.
Using this connection, we show in
section 3 that the dominant non-perturbative effects in the
quantum mechanical regularization of pure gravity
are given by a metastable instanton (bounce)
whose lifetime is universal and of the size of the non-perturbative
ambiguities in the specific heat. These results ensure, in particular,
that the method of Greensite and Halpern is compatible
with the double-scaling limit.

The problem of pure gravity and of all the even-$k$ models is that
their genus expansions for the specific heat are not Borel-summable;
hence, they are not well defined
at the non-perturbative level$^{\GM,\GZJ}$. Therefore, one should not be
surprised to have many different ``definitions'' of pure gravity if the
agreement with its genus expansion is the only selecting criterion.
In fact, a third consistent definition has been proposed recently
by imposing the non-perturbative  agreement with the KdV flow structure
of eqs. \stringeq\ and \KdV$^{\MORRIS}$.
This proposal provides a real specific
heat too, which is different from that defined by the stochastic-like
regularization$^{\MORAMB}$.
Obviously, some new physical insight into non-perturbative 2D quantum
gravity
is required to select which is the right definition. A good example
of a true physical constraint is the unitarity of the $S$-matrix in the
$d=1$ matrix model, where a clear spacetime interpretation is available,
in contrast with the $d=0$ case$^{\MOORE}$.
In fact, in $d=0$,
the agreement with the KdV flow structure could be a
good criterion, but the
matrix models provide the KdV
structure only at the perturbative level (genus by genus), and it
can be broken through non-perturbative contributions.
To conclude the paper, we discuss in section 4  the solutions of the
Hermitian and
anti-Hermitian 1-matrix models with generic potentials
that illustrate this last point.
Furthermore, we argue that the solution of the anti-Hermitian model
is the simplest way to accommodate the complex solutions of
the Painlev\'e equation that have been considered in the analytic
continuation
method, and provides another consistent definition of pure gravity.
This shows, once more, that the real problem is our
ignorance about quantum gravity at the non-perturbative level, and that
the matrix models alone do not solve these ambiguities.

\chapter{WKB APPROXIMATION AND THE SEMICLASSICAL MATRIX MODEL}

In the stochastic-like regularization method, it has been proved that
the perturbative expansion of the equivalent quantum mechanical system
is the same as that of
the original (unstable) model provided that they agree at
the first order. In the case of pure gravity,
this has already been checked using
the lowest-order potentials, where the quantum
system is an ideal Fermi gas of $N$ particles.
In this section, we show that the quantum mechanical regularization
reproduces, in the WKB approximation, the semiclassical limit of
the matrix model not only in the case of pure gravity, but for any
multicritical model.
The main difficulty is the
fact that the potentials with degree $\ge 4$ induce
interactions between the fermions,
and the Fermi gas is no longer ideal. Nevertheless, a mean field approximation
can be used to decouple the Fermi gas in the WKB approximation.

Let us consider the $d=0$ Hermitian one-matrix model, whose partition
function is
$$
Z=\int D\phi \,e^{-\beta V(\phi)} ,\eqn\partition
$$
where $\phi$ is an $N\times N$ Hermitian matrix, and $V(\phi)$ is a generic
potential of degree $L$
$$
V(\phi)  = \sum_{n=2}^{L} \,g_n \,{\rm Tr} (\phi^n) .\eqn\potential
$$
Following refs.[\QM,\KM], this model can be
formulated like the ground-state
of a $d=1$ quantum mechanical system, whose Hamiltonian is the (positive
semi-definite) Fokker--Planck Hamiltonian
$$
H_{FP} = {\rm Tr}(P^2) + W_{FP} \,\,\,{\rm  with}\,\,\,
P_{ij} = -i {\partial\over\partial
\phi_{ji} }\,\,\,{\rm and}\,\,\,
W_{FP} = {\beta^2\over4} {\partial V\over\partial \phi_{ij}}
{\partial V\over
\partial \phi_{ji}} - {\beta\over2} {\partial^2 V\over \partial \phi_{ij}
\partial \phi_{ji} }.    \eqn\HFP
$$
This means that the expectation value of a generic
operator $Q(\phi)$ in the matrix
model corresponds to the vacuum-expectation-value (VEV) of the quantum
operator $Q(\hat \phi)$:
$$
\langle Q \rangle \equiv \bra{0} Q(\hat \phi) \ket{0} = \int D\phi
\Psi_0^2(\phi) Q(\phi),\,\,\,H_{FP} \Psi_0 (\phi) = E_0\Psi_0(\phi),
\eqn\QMdef
$$
where $E_0$ and $\Psi_0(\phi)$ are the energy and the wave function of the
ground-state. $E_0 = 0$ corresponds to the case of potentials
bounded from below, whose matrix models are
well defined, and $\Psi_0(\phi) =
\exp(-\beta V(\phi)/2)$. Otherwise, if $E_0 >0$, eq.\QMdef\ defines
$\langle Q \rangle$ in terms of the true ground-state. Notice
that the extra dimension introduced is just an auxiliary degree of
freedom without interpretation within the matrix model. In fact, the
correlators of operators taken at different
times in the quantum theory do not
have any meaning in terms of the matrix model.
At this point, it is worthwhile to mention that the ground-state energy
$E_0$  is not the partition function of the $d=0$ matrix model. In fact,
this method only provides closed expressions for the correlators.
Nevertheless, it is easy to obtain directly the derivatives of the matrix model
partition function with respect to the cosmological constant, \ie, the
correlation functions of the puncture operator ${\bf \hat P}$. They are the
universal scaling parts of the correlation functions of
${\rm Tr}{\phi^n}$, for any
finite value of $n$ (see eq.(2.20)). Therefore, one could add a
perturbation to the
Fokker--Planck Hamiltonian, $H_{FP}\rightarrow H_{FP}
+J{\rm Tr}(\phi^n)$, such that
the derivative of the perturbed ground-state energy  $E_0(J)$
with respect to the source $J$, evaluated at $J=0$,
gives the correlation function of the puncture$^{\AMBA}$.

Following the well-known techniques of ref.[\BIPZ],
it is natural to make a change of variables to
the eigenvalues $\{\lambda_i\}$ of the matrix $\phi$, by
introducing the effective ground-state wave function
$$
\Psi_0^{eff} (\{\lambda_i\}) = \prod_{i<j} \left( \lambda_i - \lambda_j \right)
\Psi_0 (\phi),\eqn\antisymm
$$
which is totally antisymmetric, and describes a gas of $N$ Fermi
particles.
In general, the Fokker--Planck potential, $W_{FP}$, can be splitted into its
diagonal ($D$) and non-diagonal ($ND$) parts in terms of the
eigenvalues
$$
\eqalign{
&W_{FP}  =  W_{FP}^{(D)} + W_{FP}^{(ND)} \cr
&W_{FP}^{(D)}={\beta^2\over4} \sum_{i=1}^N
\left((V'(\lambda_i))^2 - 4X\left( g_2 + \sum_{n=3}^L
n g_n \lambda_i^{n-2}\right)\right) \cr
& W_{FP}^{(ND)} = -{\beta\over2} \sum_{i,j=1}^N
\left(\sum_{n=4}^L n g_n \sum_{s=0}^{n-4}
\lambda_i^{s+1}\lambda_j^{n-3-s} \right),\cr
} \eqn\potencialFP
$$
where $X=N/\beta=e^{\gamma_0}$ is related to the (bare) 2D cosmological
constant in the usual way. Obviously the $ND$ piece does not vanish
if $L\ge 4$, and the Fermi gas is not
decoupled. Nevertheless, in the semiclassical WKB limit
($\beta\approx \hbar^{-1}\rightarrow\infty$), a mean field
approximation (\`a la Hartree--Fock) may be performed to decouple the
system. We show below that
$$
{\rm Tr}(\phi^k) {\rm Tr}(\phi^p) \approx N \left( \omega_k {\rm Tr}(\phi^p) +
\omega_p {\rm Tr}(\phi^k)
 - N
\omega_k \omega_p \right) + \cdots  \eqn\Hartree
$$
where the normalization is fixed by
$$
\bigl\langle {\rm Tr}(\phi^k) {\rm Tr}(\phi^p)
\bigr\rangle_c \approx N^2 \left(\omega_k \omega_p +
O(1/N)\right)\,\,\,{\rm and}\,\,\,
\omega_k = {1\over N}
\bigl\langle {\rm Tr}(\phi^k)\bigr\rangle_c ,
\eqn\normaliza
$$
is consistent with the semiclassical limit of the matrix model.
Under this approximation, the quantum mechanical system
becomes an  ideal Fermi gas of $N$ particles whose Hamiltonian is
$$
\eqalign{
&H_{FP}  \approx \sum_{i=1}^N h^{FP} (\lambda_i),\,\,\,{\rm with}\,\,\,
h^{FP}(\lambda) = -
{\partial^2\over\partial\lambda^2} + {\beta^2\over4} U_{FP} (\lambda),
\,\,\,{\rm and}\cr
&U_{FP} (\lambda) = \left( V'(\lambda)\right)^2 - 4X\left( \sum_{i=1}^{L-2}
\left( \sum_{j=i+2}^L jg_j \omega_{j-i-2}\right) \lambda^i + g_2 - {1\over2}
\sum_{i=4}^L ig_i \sum_{j=2}^{i-2} \omega_{j-1} \omega_{i-j-1} \right).\cr
} \eqn\fermion
$$
The effective potential  $U_{FP}$  is bounded from below,
and the one-fermion
Hamiltonian has a well-defined discrete spectrum, $h^{FP} \psi_n(\lambda)
= {\beta^2\over4}e_n \psi_n(\lambda)$. The ground-state wave function is
the Slater determinant of the first $N$ eigenfunctions,
and the
vacuum expectation value of a generic operator $Q(\phi)$ is
$$
\bra{0} Q(\hat\phi)\ket{0}=\int [d\lambda] {\rm det}^2 \left(\psi_{i-1}
(\lambda_j)\right) Q(\{\lambda\}),\,\,\, i,j=1,\ldots,N,
\eqn\slater
$$
to be compared with the corresponding expression in the matrix model
$$
\langle Q(\phi)\rangle ={1\over Z} \int [d\lambda] {\rm det}^2 \left(
{\rm e}^{-\beta V(\lambda_j)} \lambda_{j}^{i-1}
\right) Q(\{\lambda\}),\,\,\,i,j=1,\ldots,N .
\eqn\slatMM
$$

All the relevant information
about the $d=1$ quantum mechanical system is contained in the particle density,
$\rho(\lambda,e)= \bra{\lambda} \delta \left(h^{FP} - {\beta^2\over4}e
\right)\ket{\lambda}$,
whose normalization fixes the Fermi energy  $e_F$:
$$
N = \int d\lambda\ \int_{-\infty}^{e_F} de\ \rho(\lambda,e).\eqn\fermienergy
$$
The energy integral of $\rho(\lambda,e)$
provides the quantum-mechanical version of the matrix model semiclassical
density of eigenvalues
$$
u (\lambda) = {1\over\beta} \int_{-\infty}^{e_F} de\ \rho(\lambda,e)
,\,\,\,X= {N\over\beta}=\int d\lambda\ u (\lambda),
\eqn\densitygen
$$
which, in the WKB approximation, is
$$
\eqalign{
& u^{WKB} (\lambda) = {1\over2\pi} \sqrt{e_F - U_{FP}(\lambda)}\ \theta(
e_F - U_{FP}(\lambda)) \cr
& {1\over\beta}\ \bigl\langle {\rm Tr}\ Q(\phi) \bigr\rangle =
\int d\lambda\ Q(\lambda)\,u^{WKB} (\lambda).\cr
} \eqn\densityWKB
$$

In order to compare the WKB result, eq.\densityWKB, with the matrix model,
we shall use the semiclassical limit of the Schwinger--Dyson loop
equations$^{\DAVA}$.
In the semiclassical (planar) limit, the generating function of monomial
expectation values
$$
F(p) = {1\over\beta}\,\Bigl\langle {\rm Tr}
{1\over p-\phi}\Bigr\rangle_c
\eqn\generator
$$
satisfies the Schwinger--Dyson equation
$$
F(p)^2 - V'(p) F(p) + X \sum_{i=0}^{L-2}
\left( \sum_{j=i+2}^L j g_j\,\omega_{j-i-2} \right) p^i = 0,\eqn\LOOP
$$
where the ``constants of integration''  $\omega_k$
have been already defined in eq.\normaliza. The solution of this equation is
$$
\eqalign{
&F(p) \equiv \int d\lambda\,{u^{SC}(\lambda)
\over p-\lambda} =
{1\over 2}\left( V'(p) - \sqrt{\Delta(p)}\right) \cr
& \Delta(p) = \left(V'(p)\right)^2 - 4X \sum_{i=0}^{L-2}
\left( \sum_{j=i+2}^L j g_j \omega_{j-i-2} \right) p^i,\cr
} \eqn\LOOPR
$$
and, for a given multicritical model, the ``constants of integration'' are
fixed by the condition that the imaginary part of $F(p)$ defines a proper
semiclassical density of eigenvalues$^{\DAVA,\BIPZ}$.
Obviously, $\Delta(p)$ is a polynomial in $p$ and all the
singularities of $F(p)$ will be the branch cuts of $\sqrt{\Delta(p)}$.
Therefore,
$\Delta(p)$ has to satisfy the following
constraints$^{\DAVA,\BIPZ,\Demeterfi}$:
(i) $\Delta(p)$ must have only real zeros in the complex $p$-plane,
and (ii) $\Delta(p)$ cannot have three consecutive odd degree zeros.
Under these conditions, the semiclassical density of
eigenvalues is
$$
u^{SC} (\lambda) = {1\over\pi} {\rm Im}\ F(\lambda) = {1\over2\pi}
\sqrt{-\Delta(\lambda)}\ \theta(-\Delta(\lambda)).
\eqn\densitySC
$$
The branch cuts of $F(\lambda)$, \ie, the intervals between odd
degree
(real) zeros of $\Delta(\lambda)$, are the ``bands'' on which $u^{SC}(\lambda)$
has support.\footnote{b}{All the multicritical models of refs.[\DSL,\GM]
correspond to
single-band configurations, which means that $\Delta(p)$ has exactly two
single (odd degree, in general) real zeros in the complex $p$-plane.
}

Therefore, under the above mentioned restrictions, the comparison between
eqs.\LOOPR, \densitySC\ and eqs.\fermion, \densityWKB\
shows that the WKB limit of
the $d=1$ Fokker--Planck Hamiltonian, with the mean field approximation of
eq.\Hartree, agrees with the semiclassical limit of the matrix
model if the Fermi energy is
$$
{e_F^{(L)}\over 4  X} = g_2 + 3g_3 \omega_1 + \sum_{i=4}^L
\left( \omega_{i-2} + {1\over2} \sum_{j=2}^{i-2} \left( \omega_{j-1}
\omega_{i-j-1}\right) \right) i g_i .
\eqn\Fermiomega
$$
Notice, eq.\normaliza, that the ``constants of integration'' $\omega_k$
are, in fact, correlation
functions. Accordingly, the universal piece of the Fermi energy
in the double-scaling limit can be expressed in terms of the expectation
value of the puncture
operator. In the case of even potentials, the precise relation between
$\omega_k$ and the puncture operator is, for the $k^{\rm th}$ model$^{\GM}$,
$$
N \omega_{2s} = \bigl\langle {\rm Tr} (\phi^{2s})\bigr\rangle \approx
{\rm n.u.}\,  +\,
s{2s\choose s} \beta^{-{1\over 2k +1}} \langle {\bf \hat P}\rangle ,
\eqn\puncture
$$
where n.u. stands for the non-universal part. Therefore,
the dominant universal piece of the Fermi energy is
$$
e_{F}^{(2L)} = {\rm n.u.}\ \ +\ \ \left( 4
\sum_{i=1}^{L} {(2i-2)!\over (i-1)! (i-2)!}
(2ig_{2i}) \right) \beta^{-{2k+2\over2k+1}}\, \langle{\bf\hat P}\rangle\ \
+\,\cdots
\eqn\Fermiuniv
$$
In fact, this is the most direct way to obtain the
critical exponents of the multicritical models within the quantum mechanical
formalism.
Moreover, the scaling factor when $\beta\rightarrow\infty$ will dictate the
relevant piece of the Fokker--Planck potential in the double-scaling limit.

Our result agrees with, and generalizes, previous
results obtained for pure gravity with the simplest potentials of degree
three$^{\PAR,\QM,\KM,\AMBA,\SILVB}$ and four (even)$^{\Geli}$,
showing that the critical behaviour of the
Fokker--Planck Hamiltonian is precisely that of the matrix model.
Therefore, in the semiclassical approximation, it is possible to describe
any Hermitian one-matrix model as an ideal Fermi gas of $N$ particles, whose
potential is given by the semiclassical density of eigenvalues
$$
U_{FP}(\lambda) = e_F - \lbrack 2\pi u^{SC} (\lambda)\rbrack^2 .
\eqn\potdens
$$
Notice that this relationship formally holds only when the
above mentioned constraints on
the zeros of $\Delta=U_{FP}-e_F$ are satisfied and $u^{SC}$ is well defined.
Nevertheless, the quantum mechanical system, and $u^{WKB}$, is defined even
when this is not the case. Such quantum mechanical configurations arise when
the {\it na\"\i ve} ground-state wave function,
$\Psi_0= \exp(-\beta V/2)$, is not
normalizable in the WKB approximation either.
Therefore, they do not correspond to any semiclassical configuration of the
matrix model, and they contain non-perturbative
information about the stabilization mechanism. In fact, their study
has shown that the quantum mechanical formulation is not compatible with the
KdV flows at the non-perturbative level$^{\MORAMB}$.

\chapter{UNIVERSAL NON-PERTURBATIVE BEHAVIOUR}

In the previous section, we have shown that the WKB limit of the quantum
mechanical formulation agrees with the semiclassical limit of the matrix model
for an arbitrary potential. If the stabilization mechanism is compatible with
the double-scaling limit, this result ensures that the quantum mechanical
formalism preserves the universality of the multicritical models.
In this section, we show that the non-perturbative
contributions are indeed universal in the case of pure gravity,
and that they agree qualitatively with the ambiguities expected
in the specific heat. The size of these ambiguities is obtained through the
linearization of the Painlev\'e equation, eq.\Painleve.
The result is that the
difference between any two functions whose genus expansion is
the pure gravity
genus expansion should be propotional to $T^{-{1\over 8}}
\exp\left(-{4\sqrt{6}\over5}T^{5/4}\right)$ when
$T\rightarrow\infty^{\GZJ,\SHEN}$. We
show below that the typical non-perturbative exponential suppression
corresponds to a universal mestastable instanton (bounce) in the quantum
mechanical picture, which is a natural relationship in quantum
mechanics when the perturbative expansion is not Borel-summable.

To identify the universal features of the Fokker--Planck potential,
the main tool we shall use  is the connection with the semiclassical
density of eigenvalues, eq.\potdens,
which allows the universal behaviour of the potential
to be derived from that of the semiclassical density. From now on,
we shall restrict ourselves to the case of the standard multicritical
models defined with even potentials, but the
results should be easily
extended to the general case. Using the standard orthogonal
polynomial techniques$^{\BESSIS}$, we can compute the
semiclassical density of
eigenvalues directly from the generating function of monomial
expectation values, eqs.\LOOP, \LOOPR, and \densitySC$^{\GM,\Demeterfi}$:
$$ \eqalign{
F(p)& = {1\over\beta}\, \Bigl\langle {\rm Tr} \left({1\over p -
\phi}\right)
\Bigr\rangle_c = {1\over\beta} \sum_{n=0}^{N-1} {1\over h_n^2}
\int d\lambda \, e^{-\beta V} {1\over p-\lambda}\,P_{n}^{2}(\lambda) \cr
& = {1\over\beta} \sum_{n=0}^{N-1} {1\over \sqrt{p^2 - 4 R(n)}} \rightarrow
\int_0^X dx \, {1\over \sqrt{p^2-4R(x)}}\,,\cr
} \eqn\polynomials
$$
where $R_n$ is the coefficient in the recurrence relation of the polynomials,
$\lambda P_n = P_{n+1} + R_n P_{n-1}$, $x=n/\beta$, and $R(x)$ is the limit of
$R_n$ when $\beta\rightarrow\infty$.
The imaginary part of eq.\polynomials\ provides the semiclassical density of
eigenvalues$^{\BIPZ}$
$$
u^{SC} (p) = {1\over\pi} {\rm Im} F(p) = {1\over \pi}\int_0^X dx\,
{1 \over\sqrt{4R(x)-p^2}}\,\theta( 4R(x) - p^2).\eqn\densitypol
$$
Next, we make the change of variables  $x=W(R(x))$, with $W(R)$
defined in terms of the potential as
$$
W(R)=\oint {dz\over 2\pi i} \, V'\left(z+{R\over z}\right) ,
\eqn\Wdef
$$
and the final expression for the semiclassical density is
$$
\eqalign{
u^{SC} (p) & = {1\over\pi} \int_{p^2/4}^{R(X)} dR \, {W'(R) \over
\sqrt{4R-p^2}}
\,\theta( 4R(x) - p^2)\cr
& \equiv {1\over2\pi} \sqrt{e_F -U_{FP}(p)} \,\theta( e_F -U_{FP}(p)).\cr
} \eqn\densityPOLY
$$

For the $k^{\rm th}$ model, the double-scaling
limit corresponds to $\beta\rightarrow\infty$, with\footnote{c}{As is
usual, we have fixed the critical values $X_c=R_c=1$, but
the result
is independent of this choice$^{\BARNA}$.}
$$
W(R) = 1- (1-R)^k ,\,\,\,
X = 1- \beta^{-{2k\over2k+1}} T ,\,\,\,{\rm and}\,\,\,R=  1 -
\beta^{-{2\over2k+1}} f(T).\eqn\doublesc
$$
Therefore, the critical behaviour of $u_{SC}$ is
$$
u^{SC} (p) = {2k\over \pi2^{2k}}  \int_0^{\sqrt{4R(X) - p^2}} dy \left(
4 -p^2 - y^2\right)^ {k-1}.\eqn\ufinal
$$
Notice that $U_{FP}(\lambda)$ has two roots of order $2k-1$,
$\lambda= \pm 2$, at the critical point  $X=R=1$.
Thus, the resulting Fokker--Planck potential has degree $4k-2$,
and it
corresponds to a matrix model whose potential has degree $L=2k$, \ie,
the canonical representative of the multicritical
$k^{\rm th}$ model$^{\GM}$.

The Fokker--Planck potential of pure gravity, $k=2$,
is
$$
U_{FP}(\lambda) - e_F = {1\over9}\left(\lambda^2-4R(X) \right)
\left(\lambda^2-6+2R(X)\right)^2 .
\eqn\potentialpg
$$
This potential has an absolute minimum at $\lambda=0$, and two relative minima
at $\lambda^2=6-2R(X)$; obviously, the Fermi energy is just the value of the
potential at these relative minima. The perturbative expansion around the
absolute minimum provides the WKB expansion, which reproduces
the $1/N$ expansion of the matrix model$^{\GH}$. Nevertheless, we are
interested in the double-scaling limit, which
corresponds to $\beta\rightarrow\infty$. We have already obtained the scaling
behaviour of the Fermi energy in this limit, eq.\Fermiuniv, which is like
$\beta^{-6/5}$ for pure gravity.
This behaviour corresponds to the scaling of the
potential around the secondary minima, \ie, around the Fermi energy,
which is\footnote{d}{Even though we have obtained this expression
starting from even potentials, it can also be derived from the cubic
potential$^{\KM}$, as expected from its universal character.}
$$
U_{FP}(\lambda) \approx \pm \beta^{-{6\over5}}  {16\over9}
\left( 4
z^3 - 3Tz\right) + \cdots \eqn\potfinal
$$
where $z=\beta^{2\over5}(\lambda\mp 2)$. Notice that, in this limit,
the potential becomes unbounded from below. However, this does not destroy
the stabilization mechanism. In fact, the subdominant terms in the
$\beta\rightarrow\infty$ limit are crucial to ensure that the potential has
bound-states, and, in particular, the normalization condition in
eq.\densitygen. On the other hand, the universal properties correspond to the
behaviour of the potential around the Fermi level, and  only the energy levels
close to the $N^{th}$ one are relevant in this limit. Because
$N\rightarrow\infty$, these levels do not feel the absolute minimum of the
potential, which appear to be unbounded from below. In fact, one can check that
the behaviour of the potential around $\lambda=0$ does depend on the
potential one chooses to define pure gravity, and hence it is not universal.

The dominant non-perturbative contributions are related to the classical
possibility that the
fermions at the Fermi level could be at the secondary minimum of the potential,
falling later into the main (unbounded) well
due to quantum effects (tunneling). This configuration corresponds to a
metastable instanton, also known as a bounce$^{\COL}$, which decays
through barrier penetration effects.
The lifetime of the bounce is proportional to the
imaginary part of its energy, which can be computed from
the {\it Euclidean} partition function in the semiclassical
approximation$^{\COL,\ZJBOOK}$.\footnote{e}{Notice that this imaginary
part cannot be identified with the
one obtained in the analytical continuation method, because the ground-state
energy of the quantum mechanical system is
not the specific heat of pure gravity, which is real within the
stochastic-like regularization.}
The Euclidean partition function corresponding to
the Hamiltonian of eq.\fermion,
$h^{FP}(\lambda) = -
{\partial^2\over\partial\lambda^2} + {\beta^2\over4} U_{FP} (\lambda)$,
can be written as the
$L\rightarrow\infty$ limit of the following functional integral
$$
{\rm Tr} \left( {\rm e}^{-LH}\right) =
\int_{q(-L/2)=q(+L/2)} \lbrack d q(t) \rbrack\,
{\rm e}^{- S[q(t)]} ,
\eqn\ZEuclidea
$$
where $S[q(t)]$ is the Euclidean action
$$
S[q(t)] = \int_{-L/2}^{+L/2} dt \left({1\over4} \dot q^2(t) + {\beta^2\over 4}
U_{FP}(q(t)) \right) .
\eqn\SEuclidea
$$
${\rm Tr}({\rm e}^{-LH})$ is given by $\sum\,\exp{(-LE_n)}$,
where the sum extends over the
whole spectrum. Consequently, in the $L\rightarrow\infty$ limit, the partition
function is dominated by the ground-state energy. Nevertheless, the
barrier penetration effects induce an imaginary part in the
energy value of the bounce, $E_{bc}$,
and, then, in ${\rm Tr}(e^{-LH})$. The dominant contribution can be
obtained by
computing the imaginary part of the partition function in the semiclassical
approximation. If $E_{bc}=E_{F}\, + \,i\Gamma$, where $E_{F}= {\beta^2
\over4}e_{F}$, then, in the $L\rightarrow\infty$ limit,
$$
{\rm Im}\left({\rm Tr}({\rm e}^{-LH})\right) \propto
{\rm Im}\left( {\rm e}^{-L(E_F\, +\, i\Gamma)} \right)\approx
L \Gamma\, {\rm e}^{-LE_F} ,
\eqn\Imaginary
$$
where we have taken into account the smallness of the non-perturbative
imaginary part.

In the semiclassical approximation, the dominant imaginary contribution
to ${\rm Tr}({\rm e}^{-LH})$ is given by the bounce,
which is a solution of the Euclidean equations
of motion, ${1\over2} \ddot q_c = {\beta^2\over4} U_{FP}'(q_c)$,
that starts from the secondary minimum, $q_m$,
at Euclidean {\it time}
$L=-\infty$, is reflected in the classical turning point, $q_0$,
and comes back to the minimum at {\it time}
$L=+\infty$.
Therefore, the bounce action is
$$
\eqalign{
S[q_{c}] &=
\int_{-\infty}^{+\infty} dt \left({1\over4} \dot
 q_{c}^{2}(t) +
{\beta^2\over 4} U_{FP}(q_c) \right)\cr
&=L E_F+
\int_{q_0}^{q_m} dq \sqrt{\beta^2 (U_{FP}(q) - e_F)}
\equiv L\,E_F+ S_{bc} . \cr}
\eqn\Sbounce
$$
The integration around $q_c$ in the Gaussian
approximation provides the bounce contribution to the partition function.
This calculation involves a Jacobian factor that becomes complex because of
the turning point $q_0$, and is proportional to $L$ because of translation
invariance in the Euclidean time. The final result is
$$
{\rm Im}\left( {\rm Tr}({\rm e}^{-LH})\right) \propto
L{\rm e}^{-S[q_c]}=L{\rm e}^{-LE_F -S_{bc}}\approx
{\rm Im}\left({\rm e}^{-L(E_F + i {\rm e}^{-S_{bc}})}\right) .
\eqn\Zbounce
$$
Therefore, $\Gamma\propto \exp(-S_{bc})$ is the imaginary
part of the bounce
energy, and the lifetime of the bounce is $\tau_{bc}\propto \Gamma$.
In the case of pure gravity, $S_{bc}$ remains finite in the
double-scaling limit$^{\BARNA}$, and its value is
$$
S_{bc} = {4\sqrt6\over5}  T^{5\over4} .
\eqn\finalbounce
$$
The lifetime of the bounce is exponentially
suppressed for large values of $T$, and its value is expected to give the
characteristic size of non-perturbative effects. Notice that $\exp(-S_{bc})$
is,
in fact, the size of the ambiguities between functions whose genus expansion
is that of the matrix model. Therefore, we conclude that the dominant
non-perturbative effects of the quantum mechanics regularization of the matrix
model are universal, and consistent with those expected for pure gravity. In
fact, this checks, in a non-trivial way,
that the stabilization procedure is
consistent with the double-scaling limit because
it ensures that the genus expansion coming from this method agrees with that of
the original matrix model$^{\ZJBOOK}$.

For completeness, let us consider eq.\ufinal\ for the first well-defined
multicritical model, $k=3$,
$$
u_{SC}(p)\propto \left(4R(X)-p^2\right)^{1\over2} \left(\left(p^2 + R(X) -
5\right)^2 + 5 \left(1-R(X)\right)^2\right) .
\eqn\KTRES
$$
In this case, $U_{FP}(\lambda)$ has only one absolute
minimum at $\lambda=0$. Therefore, the non-perturbative contributions to the
$k=3$ model are not related to metastable states. This agrees with the
expected relationship between metastability and non-Borel summability of the
perturbative series in quantum mechanics. Therefore, we expect that there will
be bounces in all the $k$-even models, and not in the $k$-odd
ones. This conclusion gives support to the identification of the
instabilities in the matrix model with instanton-like
contributions$^{\DAVB,\GZJ}$.
In fact, it shows that the criterion for having or not instabilities is given
by
the Fokker--Planck potential $U_{FP}$ in the double-scaling limit,
and not directly by the potential of the matrix model.

To conclude this section, let us estimate the dominant non-perturbative
contributions to the specific heat within the quantum
mechanical formalism. The matrix model partition function is
expressed in terms of the density of eigenvalues as$^{\BIPZ}$
$$
F^{MM} = \beta^2\,\int d\lambda\, u(\lambda) \left( \int d\mu\,
u(\mu)\,ln|\lambda-\mu| -V(\lambda)\right) .
\eqn\FMATRIX
$$
The dominant non-perturbative contributions to $F^{MM}$ will correspond to
the dominant non-perturbative modifications of the density of
eigenvalues, $\delta u(\lambda)$, which are related to the bounce and are
dominant outside the classical (WKB) range.
Therefore, the dominant modification of the matrix model is$^{\GZJ}$
$$
\eqalign{
\delta F^{MM} & = \beta^2 \,\int d\lambda\, \delta u(\lambda)\,\left( 2\int
d\mu\, u^{WKB}(\mu)\,ln|\lambda-\mu|\,-\, V(\lambda)\right) \cr
&=-\beta^2\, \int d\lambda \delta u(\lambda) \int_{q_0}^{\lambda} dq
\sqrt{U_{FP}(q)-e_F}\,\theta(q-q_0) ,\cr}
\eqn\MMFMOD
$$
where eq.\LOOPR\ has been used, and $q_0$ is the classical turning point at
the Fermi energy, \ie, the limit of the classically allowed range of
motion, eq.\densityPOLY.

The dominant contribution to $\delta u(\lambda)$
is related to the classical
possibility that the fermions at the Fermi energy can be sitting at the
secondary minimum. Therefore, one should expect $\delta u(\lambda)\propto
{\cal N} \delta(\lambda-q_m)$, where the normalization is expected to be
proportional to $\beta^{-1}$, because it involves only the fermions at the
Fermi energy$^{\GZJ}$, and to $\exp(-S_{bc})$, because it is
the expected suppression of the non perturbative
effects.\footnote{f}{This
last factor is apparently missing from ref.[\GZJ].}
In fact, we can check this normalization with the results of
ref.[\SILVB],
where the non-perturbative effects in the bound-state energies have been
computed, and they result to be proportional
to $\beta^{-6/5} \exp(-S_{bc})$.
Therefore, a similar contribution to the Fermi energy, $e_F$, should be
expected. This non-perturbative contribution to the Fermi energy induces a
modification in the normalization condition of the semiclassical density of
eigenvalues, eq.\densitygen, which should be compensated by the normalization
of $\delta u$
$$
X={N\over\beta} = \int {d\lambda\over2\pi} \sqrt{e_F -U_{FP}(\lambda)} + {\cal
N}.
\eqn\Newnormaliz
$$
If $e_F-e_{F}^{WKB}\propto \beta^{-6/5} \exp(-S_{bc})$,
then $\delta \left(\int
(d\lambda/2\pi) \sqrt{e_F-U_{FP}}\right)\propto
\beta^{-1} \exp(-S_{bc})$, and
the normalization of $\delta u$ behaves as expected. Therefore, we
obtain that $\delta u \propto \beta^{-1} \exp(-S_{bc})
\delta (\lambda-q_m)$, and
the dominant non-perturbative effects in the matrix model free-energy,
within the quantum mechanical formalism, are
$$
\delta F^{MM} \propto  \beta^2 {\cal N} \int_{q_0}^{q_m} dq\, \sqrt{U_{FP}
-e_F} \propto S_{bc} {\rm e}^{-S_{bc}},
\eqn\FMMMODFIN
$$
which are finite, universal, and consistent with the size
of the ambiguities of pure gravity.

\chapter{DISCUSSION AND CONCLUSSIONS}

As we said in the introduction, we should expect to have many
different models with the same genus expansion as
pure gravity or any other $k$-even multicritical model,
and it is necessary to have a criterion to decide which one is describing
2D quantum gravity at the non-perturbative level. It seems very
difficult to obtain such a requirement from the matrix model,
where all the results are linked to the double-scaling limit.
In fact, even the KdV structure, which is a very strong and
important result,
holds only genus by genus, and the matrix models themselves
provide examples
where it is broken at the non-perturbative level.
In particular, we shall briefly discuss
the solutions of the Hermitian
and anti-Hermitian one-matrix models with generic potentials, where
this is the case.

The solution of the Hermitian one-matrix models with generic potential,
in the one-arc sector, is well-known$^{\PETRO}$. There exist two
functions, $f^{(\pm)}_{H}=f_{H}\pm g_{H}$, that satisfy
the string equation, eq.\stringeq, and in terms of which the KdV flows can
be constructed. Their sum, $f_{H}$, is the specific heat of the
matrix model, and $g_{H}$ is a non-perturbative function whose
normalization should be related
to the couplings of the odd powers of the matrix in the potential.
Therefore, the KdV flow structure is not realized in terms of the specific
heat because of non-perturbative contributions. In the particular case of pure
gravity, $k=2$, the two functions $f^{(\pm)}_{H}$ satisfy the Painlev\'e~I
equation, eq.\Painleve, which can be written in terms of the specific heat as
$$
\eqalign{
&0\,=\,2f_{H}g_{H}\,-\,{1\over3}\,g_{H}''\cr
&T\,=\,f^{2}_{H}\,-\,{1\over3}\,f''_{H}\,+\,g^{2}_{H}.\cr}
\eqn\NEWA
$$
The dominant behaviour  for large positive values of $T$ is
$f_{H}\approx\sqrt{T}$, as required, and $g_{H}\propto T^{-{1\over8}}
\exp(-{4\sqrt6\over5} T^{5/4})$.
Therefore, the specific heat satisfies an equation
different from the Painlev\'e~I
equation, and the difference is non-perturbative. Of course, this provides an
additional definition of pure gravity to those described in the introduction,
but we are again restricted to the real
solutions of the Painlev\'e~I equation, and the problems pointed out by
David in ref.[\DAVA] remain.

Another interesting example is the solution of the
anti-Hermitian models$^{\TIM}$. In this case, also in the one-arc sector with
a generic potential, there is one complex function, $\chi_{A}=f_{A}+ig_{A}$
satisfying the string equation, and in terms of which the
KdV flows can be expressed. The specific heat
is given by the real part of this function, and it is obviously real, while
the imaginary part is non-perturbative. The
correlation functions are given by the real part of the KdV flow equations
$$
{\partial  f_{A}\over\partial t_k} \equiv \langle {\bf \hat P \hat P} {\bf
\hat O_k} \rangle = Re\left(R_{k}'\right).
\eqn\KdVcomplex
$$
Therefore, again, the KdV flow structure is not realized in terms of the
specific heat because of non-perturbative contributions. In particular, the
equations defining pure gravity are now
$$
\eqalign{
&0\,=\,2f_{A}g_{A}\,-\,{1\over3}\,g_{A}''\cr
&T\,=\,f^{2}_{A}\,-\,{1\over3}\,f_{A}''\,-\,g^{2}_{A}.\cr}
\eqn\NEWB
$$

The definition of pure gravity from the anti-Hermitian matrix model has  some
nice relevant features. It was shown in ref.[\DAVA] that no real solution
of the Painlev\'e~I equation can be consistent with the loop equations of the
matrix model, because all of them have poles
along the real axis. Nevertheless,
this is not the case with the complex solutions, and there is a unique one
without poles along the real axis and consistent with the
required asymptotic behaviour of the specific heat, $f_{A}\approx\sqrt{T}$
when $T\rightarrow+\infty$: the
``triply truncated solution''$^{\DAVA,\DAVB}$.
In contrast with the case of Hermitian matrix models, the complex solutions of
the Painlev\'e equation are natural in the context of the
anti-Hermitian models. Therefore,
this definition of pure gravity using the ``triply truncated solution'' is
expected to be consistent with the matrix model loop equations, and with
the reality of the specific heat. It is also worth noticing the
similarity of this solution to the one proposed in ref.[\MORRIS]
(compare the ``Re f'' in fig.(1) of ref.[\SILVA] with fig.(1) of
ref.[\MORRIS]). In fact, their asymptotic behaviours are the same
also for the large negative values of $T$, where $f_{A}\approx0$,
and it could be interesting to investigate their relationship further.

To summarize, we have shown that the stochastic, quantum, or $d=1$-like
definition of the $d=0$ matrix models preserves universality in the
double-scaling limit. Moreover, the dominant non-perturbative
contributions of the even-$k$ models are expected to be related to
the existence of bounces, showing their instability.
Nevertheless, we would like to finish by just repeating
that the essential
problem, from the 2D gravity point of view,
is how to choose between all the possible definitions compatible with
the genus expansion of the matrix models,
and that some new information about the non-perturbative behaviour
is needed to solve it.
\line{}
\line{}
\line{}
\centerline{\bf ACKNOWLEDGEMENTS}
\noindent
It is a pleasure to thank Enrique Alvarez and Juan Ma\~nes for a critical
reading of this manuscript.
JSG thanks also Francis Halzen for his hospitality at
the Department of Physics
of the University of Wisconsin at Madison, where part of this work has
been done. This work has been partially supported by CICYT (Spain).
\endpage
\refout
\end